\documentclass[sigconf]{acmart}

\usepackage{balance}

\def\BibTeX{{\rm B\kern-.05em{\sc i\kern-.025em b}\kern-.08emT\kern-.1667em\lower.7ex\hbox{E}\kern-.125emX}}
    
\setcopyright{acmcopyright}

\begin{document}

\fancyhead{}

\title{Interactive Variance Attention based Online Spoiler Detection for Time-Sync Comments}

\author{Wenmian Yang}
\orcid{0000-0001-8493-4449}
\affiliation{%
  \institution{Shanghai Jiao Tong University}
    \institution{State Key Lab of IoT for Smart City, CIS, University of Macau}
}
\email{sdq11111@sjtu.edu.cn}

\author{Weijia Jia}
\authornote{Corresponding author}
\affiliation{%
    \institution{State Key Lab of IoT for Smart City, CIS, University of Macau}
  \institution{Shanghai Jiao Tong University}
  }
\email{jiawj@um.edu.mo}

\author{Wenyuan Gao}
\affiliation{%
  \institution{Shanghai Jiao Tong University}
  }
 \email{gaowenyuan1028@sjtu.edu.cn}
 
\author{Xiaojie Zhou}
\affiliation{%
  \institution{Shanghai Jiao Tong University}
  }
  \email{ szxjzhou@sjtu.edu.cn}

\author{Yutao Luo}
\affiliation{%
  \institution{Shanghai Jiao Tong University}
  }
\email{luoyt1996@sjtu.edu.cn}

%
\renewcommand{\shortauthors}{Wenmian Yang, et al.}

%
\begin{abstract}
Nowadays, time-sync comment (TSC), a new form of interactive comments, has become increasingly popular on Chinese video websites. By posting TSCs, people can easily express their feelings and exchange their opinions with others when watching online videos. However, some spoilers appear among the TSCs. These spoilers reveal crucial plots in videos that ruin people's surprise when they first watch the video. In this paper, we proposed a novel Similarity-Based Network with Interactive Variance Attention (SBN-IVA) to classify comments as spoilers or not. In this framework, we firstly extract textual features of TSCs through the word-level attentive encoder. We design Similarity-Based Network (SBN) to acquire neighbor and keyframe similarity according to semantic similarity and timestamps of TSCs. Then, we implement Interactive Variance Attention (IVA) to eliminate the impact of noise comments. Finally, we obtain the likelihood of spoiler based on the difference between the neighbor and keyframe similarity.  Experiments show SBN-IVA is on average 11.2\% higher than the state-of-the-art method on F1-score in baselines.
\end{abstract}

\copyrightyear{2019} 
\acmYear{2019} 
\acmConference[CIKM '19]{The 28th ACM International Conference on Information and Knowledge Management}{November 3--7, 2019}{Beijing, China}
\acmBooktitle{The 28th ACM International Conference on Information and Knowledge Management (CIKM '19), November 3--7, 2019, Beijing, China}
\acmPrice{15.00}
\acmDOI{10.1145/3357384.3357872}
\acmISBN{978-1-4503-6976-3/19/11}

%
%



%

%
\maketitle

\section{Introduction}
\label{intro}
Recently, people enjoy watching videos online in their leisure time. During viewing time, many people are willing to share their feelings and exchange ideas with others. Time-sync comment (TSC) is such a new form of interactive comment designed to meet this demand. People can post TSCs at any time during their viewing, and these comments posted will be seen by other viewers synchronously on the video screen. However, many TSCs contain the spoiled content that reveal crucial plots in videos such as the content of movie ending, the murderer identity of criminal investigation TV-series, or final scores of sports games which can ruin the surprise of those who have not watched this video. To avoid being spoiled, lots of people have no choice but to hide all the TSCs regardless of the TSCs are spoilers or not. Some people even stay away from certain online video websites with TSCs to minimize the chances of being spoiled, which makes the users lose the opportunity to discuss with others. To ensure that users can communicate through TSCs and avoid being spoiled by TSCs, it is critical to detect spoilers from TSCs.

Most of the existing spoiler detection methods are based on key-matching and traditional machine learning methods. For instance, Nakamura \emph{et al.} \cite{Nakamura07} and Golbeck \emph{et al.} \cite{Golbeck12} filter out spoilers according to predefined keywords. Guo \emph{et al.} \cite{Guo10}  use LDA-based machine learning method to rank the most likely spoilers.  Chang \emph{et al.} \cite{chang2018deep} propose a deep neural spoiler detection model using a genre-aware attention mechanism. However, these methods are based on common reviews or articles on social media. They are not well designed for TSC data since they do not consider the unique properties of TSCs. Distinguished from common video comments, TSCs have short-text, interactive, real-time and high-noise properties.

\begin{figure}[htbp]
  \centering
  \includegraphics[width=1\linewidth]{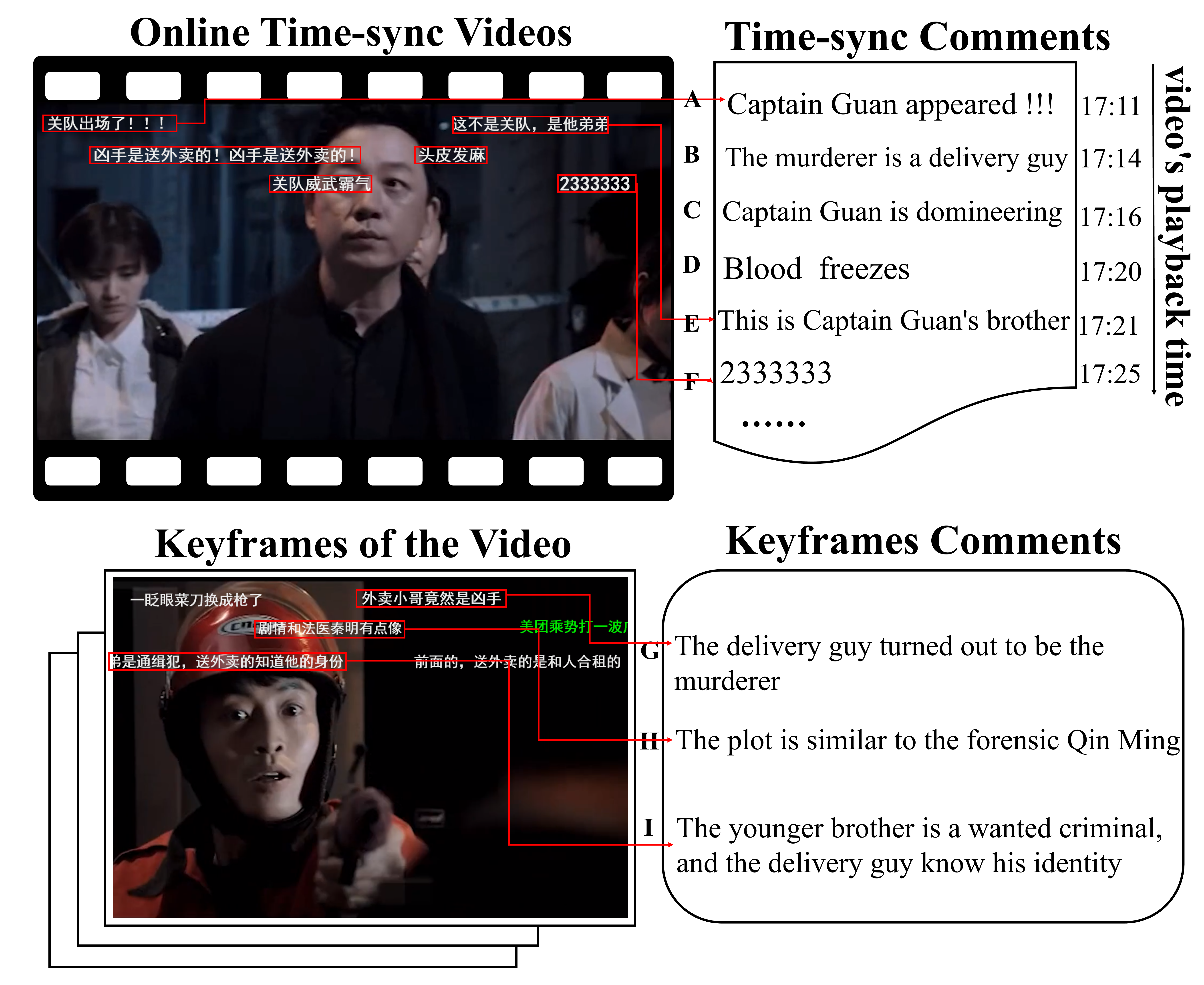}
  \caption{Example of the short-text, interactive, real-time and high-noise properties of TSCs. }\label{fig:tscintro}
\end{figure}

The most prominent property of TSCs is that they are \textbf{short texts} with lots of network slang. TSCs are usually not complete sentences, and generally, have no more than 10 words \cite{Wu14}. The words in TSCs do not contribute equally to determining the spoilers. An example of criminal investigation TV-series is shown in Fig. \ref{fig:tscintro}. The words "murderer", "delivery guy" and "Captain Guan" are plot-related words or important character names, which have a higher possibility to trigger spoiled contents. Therefore, we need to pay different attention to words with different importance in TSCs. 

Another important property of TSCs is \textbf{interactive}. The topic of the latter posted TSCs usually depends on the former, which is similar to the herding mentality effect \cite{He16,Yang19,yang2019time} in social science. In Fig. \ref{fig:tscintro}, the TSC A, C, and E share the same topic of discussing Captain Guan when Captain Guan's face appears on the screen. TSC A first shows his interest in the appearance of Captain Guan. Then this topic arouses other viewers' interests, and other TSCs within the same topic (TSC C and E) appear due to the impact of TSC A. In the example,  the generation of a TSC is not independent but influenced by the previous TSCs. This interactive property is helpful to distinguish spoiler TSCs: If the topic of a TSC is closed to its surrounding TSCs, it is less likely to be a spoiler because it describes the current content in the video.

The \textbf{real-time} property is also helpful to study the relationship between TSCs. As shown in Fig. \ref{fig:tscintro}, each TSC has a timestamp synchronous to the playback time of the video, and its content is often related to what is happening currently in the video. Therefore, a TSC is more likely to become a spoiler when it has a high semantic similarity to the TSCs of the video highlights (such as the end of the movie, the moment of revealing the murderer in the drama, and the final stage of the sports game). In this paper, we call those video highlights as ``keyframes''. In Fig. \ref{fig:tscintro}, the content of TSC B is different from its surrounding TSCs but very close to the keyframe TSC G, which means TSC B describes an important plot on purpose and has high possibility to be a spoiler. And actually, TSC B is a spoiler indicating the identity of the murderer. 

In addition to the above three properties, TSCs are \textbf{high-noise}. Many TSCs unrelated to the video content appear due to the viewers' diverse topics and particular ideas. For example, TSC D and F in Fig. \ref{fig:tscintro} express viewers' feelings irrelevant to the video content. They have weak semantic relevance with their neighboring TSCs, and they are not spoilers, either. The existence of noise causes great troubles to establish semantic relationships between TSCs. Therefore, how to make use of the interactive and real-time properties of TSCs and eliminate the impact of noise is the central challenge to predict the spoilers accurately and effectively.

Based on the above motivation and challenges, we propose a novel Similarity-Base Network with Interactive Variance Attention (SBN-IVA) to classify TSCs as spoilers or not. Specifically, due to the short-text property, we first extract textual features of TSCs through the word-level attentive encoder, which assigns the weights based on the importance of words. Then, based on the interactive and real-time properties of TSCs, we design a Similarity-Base Network (SBN) to acquire the similarities between each TSC and its neighbors and keyframes respectively according to the semantic and timestamps. The higher the neighbor similarity and the lower the keyframe similarity, the less likely the TSC is a spoiler. Considering the high-noise property, we implement Interactive Variance Attention (IVA) mechanism in the framework to effectively reduce the impact of noise due to the low semantic similarities between the noise and its surrounding TSCs \cite{Yang17}. Finally, we obtain the likelihood of the spoilers based on the difference between the neighbor and keyframe similarity.

The main contributions of this paper are as follows:
\begin{itemize}
\item[1] We propose a novel Similarity-Base Network with Interactive Variance Attention (SBN-IVA) to detect spoilers from TSCs. 

\item[2] We combine attention mechanism both in word-level and sentence invariance level in our model, which takes the short-text, interactive and real-time properties of TSCs into account and effectively reduces the impact of noise.
\item[3] We evaluate our proposed model with real-world datasets in the categories of movies, TV-series, and sports on mainstream video websites. Experiments show our model outperforms the baselines in Precision, Recall, and F1-Score.
\end{itemize}

\section{Related Work}

In this section, we introduce the related work in three aspects.

\textbf{Time-sync comment (TSC)} provides a new data source for text mining regarding the online videos and attracts research on it increasingly.  Wu \emph{et al.} \cite{Wu14} first formally introduce the TSC, and propose a preliminary method to extract video tags. Yang \emph{et al.} \cite{Yang17} systematically collate the features of TSCs and introduces a graph-based algorithm to eliminate the impact of noise prominently. Then further usages of TSCs data are proposed like extracting highlight shots \cite{Xian15,ping2017video},  labeling important segments \cite{Lv16}, detecting events \cite{li2016event}, generating temporal descriptions of videos \cite{Xu17}, and video recommendation \cite{Yang19,chen2017personalized,ping2018video}. Besides, some other researchers turn to build and improve TSC dataset. Liao \emph{et al.} \cite{liao2018tscset} present a rich self-labeled TSC dataset with four-level structures for user experience improvement. Chen \emph{et al.} \cite{chen2018fine} and Chen \emph{et al.} \cite{Chen17} also do similar evaluating work on real-world multimedia dataset. The above methods show the promising potential of TSC data and greatly inspire our work.

\textbf{Spoiler detection} (or plot detection) has achieved great attention on studies for review documents or articles in social media. The recent studies of spoiler detection mainly focus on two domains, keyword matching, and machine learning methods. Keyword matching methods filter out spoilers based on predefined keywords, such as actor and character names \cite{Golbeck12}, the name of the sports team or sporting event\cite{Nakamura07}, or words in the latter half of the document \cite{Kyosuke16}. However, the keyword matching methods require human-fixed input and are not widely used in various application scenarios. Besides, keyword matching methods usually have high recall and low precision performance since they treat many positive comments as spoilers. The other domain of research is machine learning methods. Guo \emph{et al.} \cite{Guo10} use bag-of-words representation and LDA-based model to rank spoilers. They calculate the spoiler probability scores by the similarity between comments and item descriptions from IMDB. Iwai \emph{et al.} \cite{Iwai14} evaluate five conventional machine-learning methods and improve the method through generalizing keywords. Jeon \emph{et al.} \cite{Jeon13} detect spoilers by SVM classification according to four features: "named entity", "frequently used verb", "objectivity + URL", and "tense". Hijikata \emph{et al.} \cite{Yoshinori16} introduce location information and neighborhood plot probability to help identify spoilers with contextual information by SVM. Moreover, Chang \emph{et al.} \cite{chang2018deep} propose a deep neural spoiler detection model using a genre-aware attention mechanism. However, the above methods are not well-designed for detecting spoilers in TSC data, as they ignore the interactive, real-time, and high-noise properties of TSC data.

\textbf{Attention Mechanism} is firstly proposed and used in machine translation \cite{Bahdanau14,bojar2017findings} in natural language processing. They integrate attention mechanism with an encoder-decoder framework to align the phrase in source language with the target language before translation. Vinyals \emph{et al.} \cite{Vinyals14} introduce a novel agnostic attention-enhanced seq-to-seq model to solve syntactic constituency parsing problem. Sukhbaatar \emph{et al.} \cite{Sukhbaatar15} and Kumar \emph{et al.}\cite{Kumar15} use the attention mechanism to improve traditional methods in question answering problem. Recently, Vaswani \emph{et al.} propose a new kind of seq-to-seq network architecture, Transformer \cite{Vaswani17,devlin2018bert,shaw2018self,you2019improving}, which is more suitable to solve the long-range dependencies problems. However, since the TSCs are usually short-text, the Transformer is not suitable for TSCs. Our work adopts attention mechanisms both in word and sentence level. The word-level attentive encoder assigns the high weight to informative words as Sukhbaatar \emph{et al.} \cite{Sukhbaatar15}. The sentence-level Interactive Variance Attention (IVA) refers to soft attention \cite{Luong15,yao2015video, yang2019legal}, where higher attentive weights indicate that the corresponding features are more informative for the end task.

\section{PRELIMINARIES AND PROBLEM DEFINITION}
In this section, we first provide the problem definition in Section \ref{pd}. And then, the problem formulation is introduced in Section \ref{pf}.
\begin{table}[htbp]
  \centering
  \caption{Notations and descriptions.}
  \label{tab:Notations}
  \begin{tabular}{|c|c|}
    \toprule
    Notations&Descriptions\\
    \midrule
    $\boldsymbol{TSC}$ & TSC sequence\\
    \hline
    $TSC_i$ &$i$-th TSC in $\boldsymbol{TSC}$\\
    \hline
    $L$ & Number of TSCs.\\
    \hline
    $\boldsymbol{T}$ & Timestamps sequence of TSCs\\
    \hline
    $t_i$ & Timestamp of $TSC_i$\\
    \hline
    $\boldsymbol{WL_{i}}$ & Word sequence of $TSC_{i}$\\
    \hline
    $K$ &Number of words\\
    \hline
    $wl_{i,j}$ &$j$-th word of $TSC_i$\\
    \hline
    $R$ & Number of former TSCs\\
    \hline
    $\boldsymbol{F}$ & Sequence of former neighbors\\
    \hline
    $F_{i}$ & $i$-th former neighbor\\
    \hline
    $P$ & Number of keyframes\\
    \hline
    $\boldsymbol{KEY}$ & Sequence of keyframes\\
    \hline
        $KEY_{i}$ & $i$-th keyframe\\
\hline
$\boldsymbol{x_{i,j}}$ & Embedding vector of $w_{i,j}$\\
\hline
$LSTM$ & Long Short Term Memory Network\\
\hline
$\boldsymbol{h_{i,j}}$ & LSTM hidden state of $\boldsymbol{x_{i,j}}$\\
\hline
$\alpha_{i,k}$ & Word-level attention score of $\boldsymbol{h_{i,j}}$\\
\hline
$\boldsymbol{W_{s}}$ & Weight matrix\\
\hline
$\boldsymbol{u_{s}}$ & Context vector\\
\hline
$\boldsymbol{Tseq_{i}}$ & Target sentence vector of $TSC_{i}$\\
\hline
$\boldsymbol{NSEQ}$ &  Sentence vector sequence of former neighbors\\
\hline
$\boldsymbol{Nseq_{i}}$ &  Sentence vector of former neighbor $F_{i}$\\
\hline
$\boldsymbol{KSEQ}$ &  Sentence vector sequence of keyframes\\
\hline
$\boldsymbol{Kseq_{i}}$ &  Sentence vector of keyframe $KEY_{i}$\\
\hline
$Nsim_{r}$ & Semantic similarity between $\boldsymbol{Tseq_{i}}$ and $\boldsymbol{Nseq_{r}}$\\
\hline
$G_{Nsim_{i}}$ & Weighted average similarity of $TSC_i$'s neighbors\\
\hline
$\beta$ & Hyper-parameter of decay function\\
\hline
$Ksim_p$ & Semantic similarity between $\boldsymbol{Tseq_{i}}$ and $\boldsymbol{Keq_{q}}$\\
\hline
$G_{Ksim_{i}}$ & Maximum keyframe similarity of $TSC_i$\\
\hline
$\hat{y_i}$ &  Prediction result of $TSC_i$\\
\hline
$y_i$ &  Ground truth of $TSC_i$\\
\hline
$\boldsymbol{S_{r}}$ & Semantic similarity of $F_r$ to other neighbors\\ 
\hline
$S_{i,j}$ & Semantic similarity between $\boldsymbol{Nseq_{i}}$ and  $\boldsymbol{Nseq_{j}}$\\
\hline
 $\boldsymbol{\overline S_{r}}$ & Normalized similarity of $F_r$ to other neighbors \\
\hline
 $\overline S_{i,j}$ & Normalized similarity between $\boldsymbol{Nseq_{i}}$ and  $\boldsymbol{Nseq_{j}}$\\
\hline
$D_r$ & Similarity variance of  $\boldsymbol{\overline S_{r}}$\\
\hline
$\overline D_{r}$ & Sentence-level attention score of $F_r$\\
\hline
  \bottomrule
\end{tabular}
\end{table}

\subsection{Problem Definition}
\label{pd}
In this paper, we aim at detecting and labeling spoilers, i.e., TSCs containing certain plots in videos. We call the TSC we are studying as the target TSC. The key idea of our framework is to compare the neighbor similarity and the keyframe similarity of the target TSC. Intuitively, we build a similarity-based deep framework to solve this task in a supervised way. Specifically, we first represent the TSCs as semantic vectors of through word-level attentive encoder. Then, we compute the neighbor similarity and the keyframe similarity of the target TSC though SBN with sentence-level IVA. The definitions of the neighbor similarity and the keyframe similarity in our two-level framework are shown as follows:

\textbf{Neighbor similarity:}
Former neighbors are several consecutive TSCs in front of the target TSC. We define the semantic similarity between target TSC and each of its former neighbors as neighbor similarity. We employ the weighted average of the neighbor similarities as overall neighbor similarity for each target TSC. The high overall neighbor similarity means that the target TSC describes current video content, and thus it is less likely to be a spoiler.

\begin{figure}[htbp]
  \centering
  \includegraphics[width=1\linewidth]{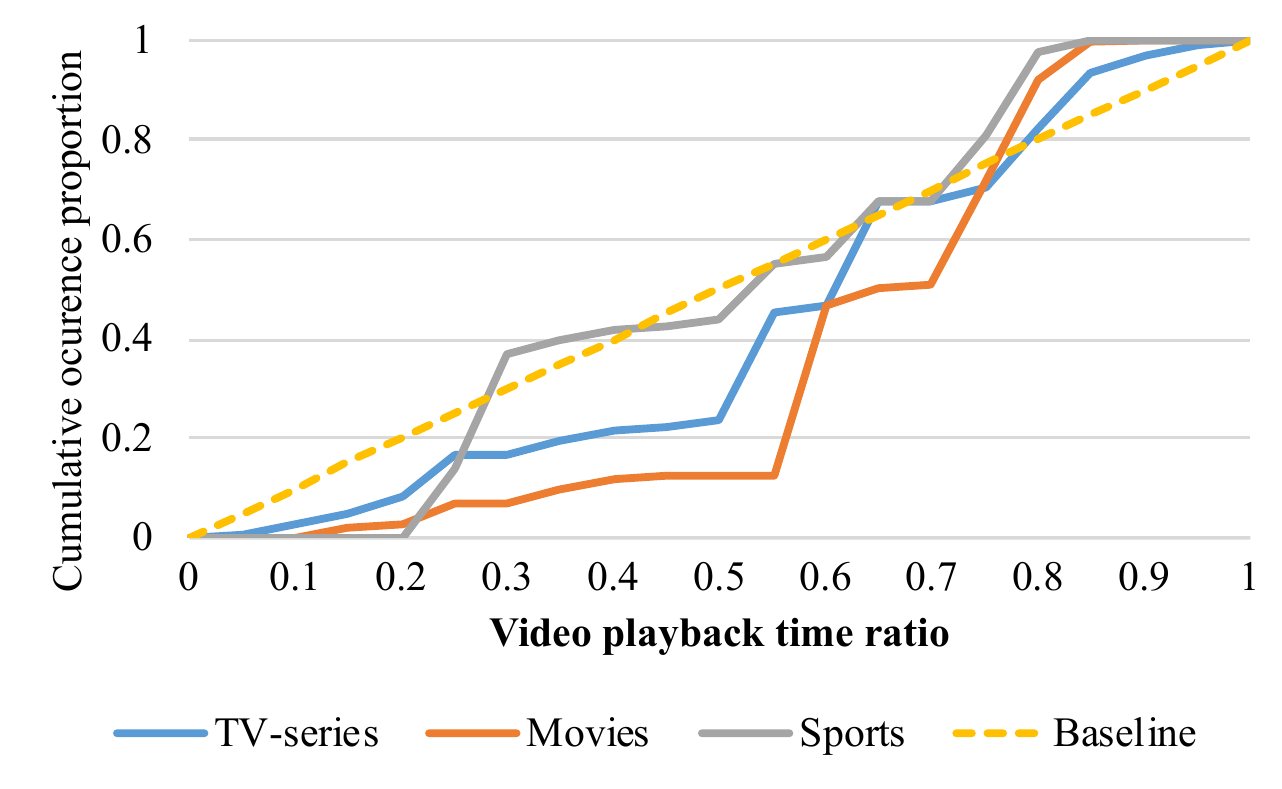}
  \caption{Cumulative occurrence proportion of key-TSCs with the video's playback time ratio.}\label{fig:keyframe}
\end{figure}

\textbf{Keyframe similarity:}
Keyframes are the highlight clips of the videos, such as the end of the movie, the moment of revealing the murderer in the drama, and the final stage of the sports game. The TSCs in the keyframes contain a large number of keywords that are related to the videos' important plots. If these keywords appear in the TSCs before these keyframes, these TSCs are spoilers. For each video, we define the similarity between target TSC and each keyframe as keyframe similarity. If the content of the target comment is similar to any of the keyframe, it is very likely to be a spoiler. Therefore,  we take the maximum keyframe similarity as the overall keyframe similarity. We perform a qualitative analysis of the cumulative occurrence proportion of key-TSCs (TSCs with keywords) with the video's playback time ratio. As shown in Fig. \ref{fig:keyframe}, the key-TSCs usually post in the last quarter of the video. Therefore, the keyframes generally appear in the last quarter of the video. Moreover, the keyframes are usually the most intensive time periods of TSCs \cite{Xian15}.  As a result, we cut the last quarter of the video into frames,  and the duration of each frame is 10 seconds. We count the number of TSCs contained in each frame and select P frames with the most TSCs as keyframes.

\subsection{Problem Formulation}
\label{pf}
For each video, we have the TSC sequence $\boldsymbol{TSC}$ = $\{TSC_{1},TSC_{2},...,$ $TSC_{L}\}$
, where $L$ is the number of TSCs in the video. We define $ \boldsymbol{T} = \{t_{1},t_{2},...,t_{L}\}$ as the corresponding timestamps of TSCs, where $t_{1} \leq t_{2} \leq ... \leq t_{L}$. The word list of $TSC_{i}, i \in[1,L]$ is defined as $\boldsymbol{WL_{i}} = \{wl_{i,1},wl_{i,2},...,wl_{i,K}\}$, where $K$ is the number of words in $TSC_i$.

For each target $TSC_{i}$, we define its $R$ former TSCs as former neighbors $\{TSC_{i-R},TSC_{i-R+1}, ... ,TSC_{i-1}\}$. The target $TSC_{i}$ and its $R$ former neighbors are consecutive in the TSC sequence of each video, so their timestamps $t_{i-R}<t_{i-R+1}<...<t_{i}$. To simplify the annotation, we use $\boldsymbol{F} = \{F_{1},F_{2}, ... ,F_{R}\}$ to indicate the set of former neighbors, and $\{t_{F_{1}},t_{F_{2}}, ... ,t_{F_{R}}\}$ to indicate their corresponding timestamps, where $t_{F_{1}} \leq t_{F_{2}} \leq ... \leq t_{F_{R}}$.  

We also give the definition of keyframes of each video as $\boldsymbol{KEY} = \{KEY_{1}, KEY_{2},...,KEY_{P}\}$, where $P$ is the number of keyframes in each video. As mentioned in Section 3.2, we treat the $P$ most intensive periods of TSCs in the last quarter of each video as the keyframes.

Given all the $ \boldsymbol{TSC} = \{TSC_{i}|  1\leq i\leq L\}$ and their corresponding timestampes $ \boldsymbol{T} = \{t_{1},t_{2},...,t_{L}\}$, our work is to predict whether $TSC_{i}$ is a spoiler or not. To make more clear presentation, we list the notations used throughout the paper in Table~\ref{tab:Notations}.

\section{Model}
In this section, we first use the word-level attentive encoder to extract textual features of TSCs in Section \ref{4.1}. Then, we propose a novel Similarity-Based Network to detect the spoilers in Section \ref{4.2}. To further make use of the interactive and real-time properties and eliminate the impact of noise among TSCs, we implement Sentence-Level Interactive Variance Attention(IVA) to improve the detection accuracy of SBN in Section \ref{4.3}.

\subsection{Word-Level Attentive Encoder}
\label{4.1}
\begin{figure}[htbp]
  \centering
  \includegraphics[width=1\linewidth]{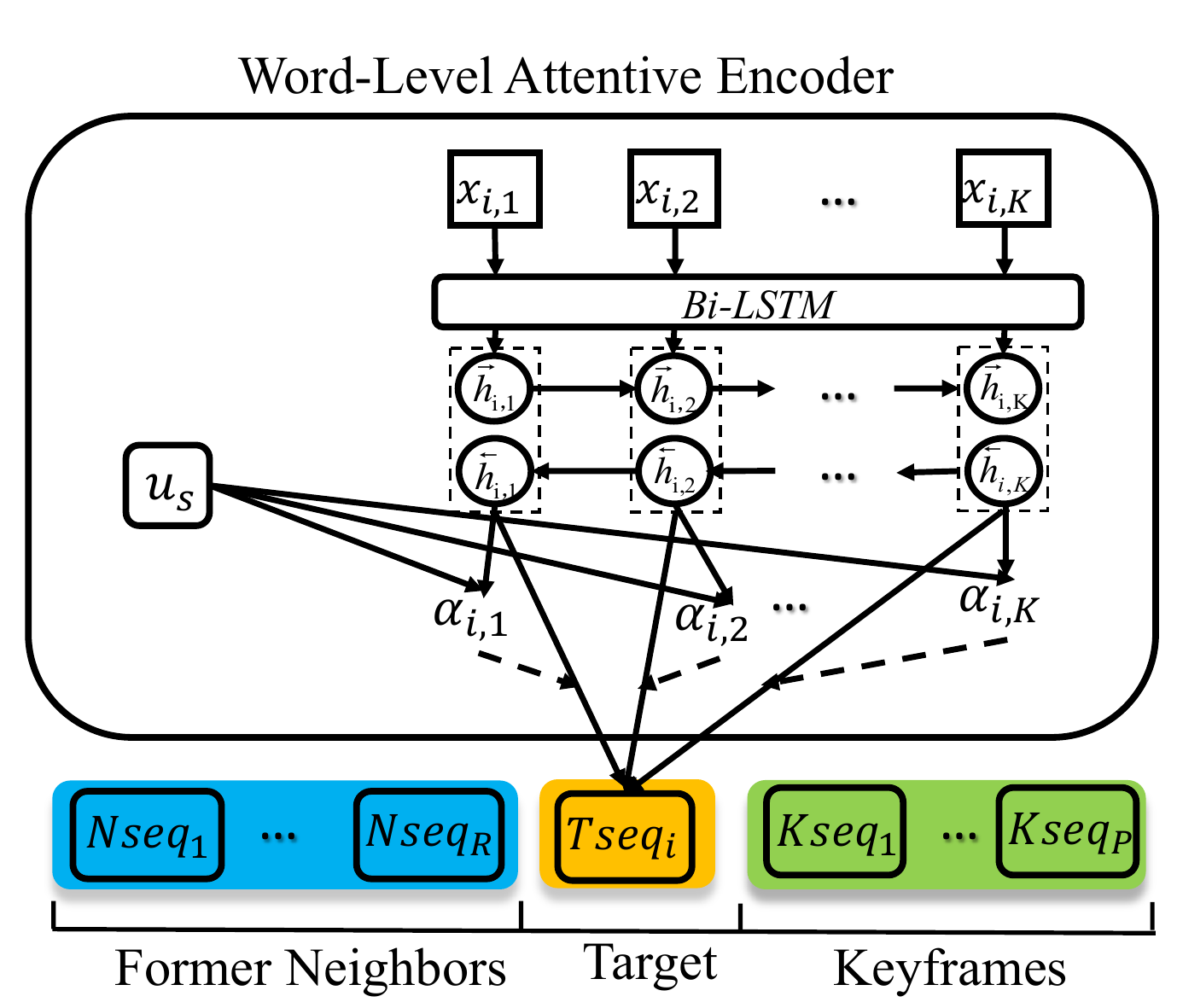}
  \caption{Word-level attentive encoder of target TSC and its former neighbors.}\label{fig:wordencoder}
\end{figure}

\begin{figure*}[htbp]
  \centering
  \includegraphics[width=1\linewidth]{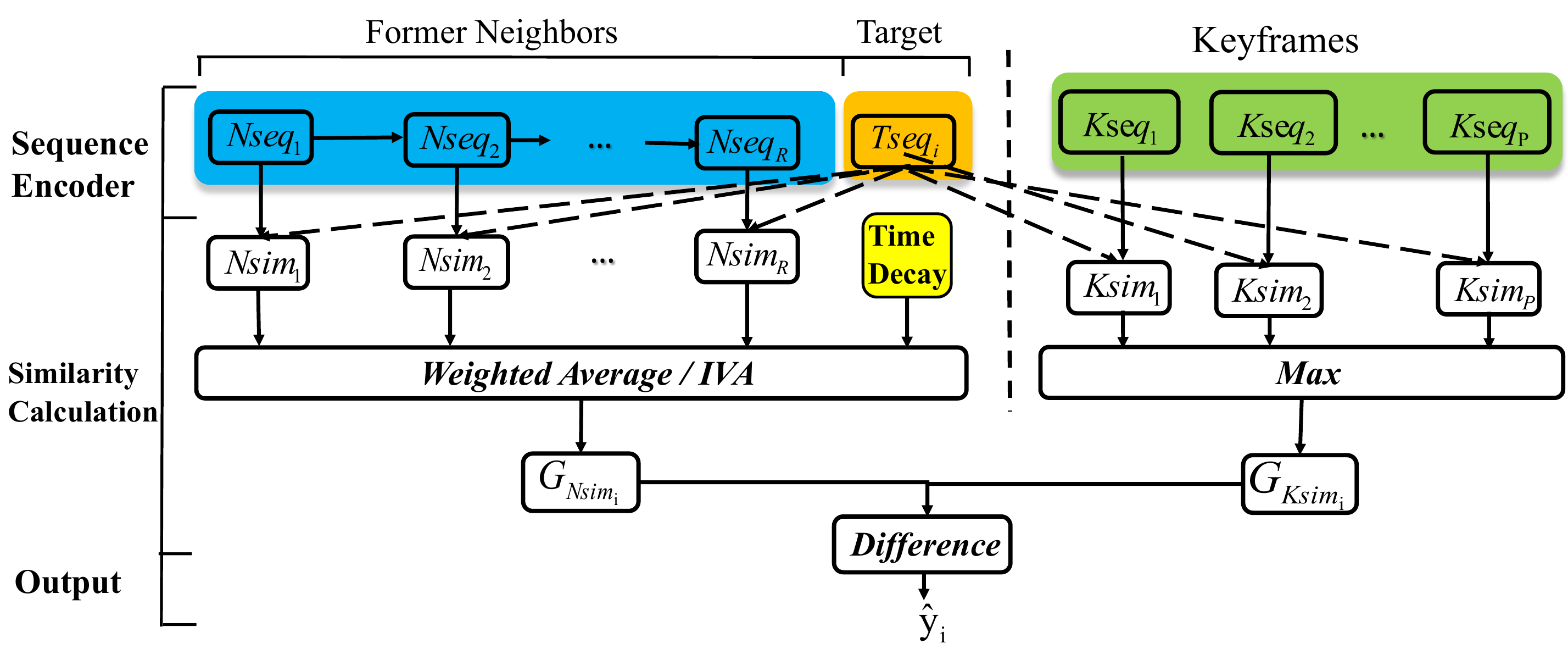}
  \caption{Similarity-Based Network.}\label{fig:seq}
\end{figure*}
In this section, we use the Bidirectional Long Short-Term Memory (Bi-LSTM) based word-level attentive encoder to extract textual features of each TSC. The architecture of the word-level attentive encoder is shown in Fig. \ref{fig:wordencoder}. 

Given the word sequence $\boldsymbol{WL_{i}} = \{wl_{i,1}, wl_{i,2},$ $..., wl_{i,K}\}$ of the $TSC_{i}$, we first embed the word $wl_{i,k}$ into fix-sized vector $\boldsymbol{x_{i,k}}$. We pre-train the embedding matrix  by Skip-Gram model \cite{Mikolov2013Efficient} because it is better for infrequent words that often appear in TSCs. Afterwards, we use Bi-LSTM network \cite{Hochreiter97}  to extract textual features of each TSC. Specifically, for each $\boldsymbol{WL_{i}} = \{wl_{i,1},wl_{i,2},...,wl_{i,K}\}$, we have
\begin{equation}
 \boldsymbol{ \overrightarrow{h} _{i,1}}= \overrightarrow{LSTM}( \boldsymbol{x_{i,1}})
\end{equation}
\begin{equation}
 \boldsymbol{ \overleftarrow{h} _{i,K}}= \overleftarrow{LSTM}( \boldsymbol{x_{i,K}})
\end{equation}
\begin{equation}
 \boldsymbol{ \overrightarrow{h} _{i,k}}= LSTM(  \boldsymbol{{x_{i,k}}, \boldsymbol{ \overrightarrow{h} _{i,k-1}})}, \boldsymbol{\overrightarrow{h} _{i,k-1}}, k \in [2,K]
\end{equation}
\begin{equation}
 \boldsymbol{ \overleftarrow{h} _{i,k}}= LSTM ( \boldsymbol{{x_{i,k}}, \boldsymbol{ \overleftarrow{h} _{i,k+1}})}, \boldsymbol{\overleftarrow{h} _{i,k+1}}, k \in [1,K-1]
\end{equation}
and
\begin{equation}
\boldsymbol{h_{i,k}}= (\boldsymbol{\overrightarrow{h} _{i,k}} || \boldsymbol{\overleftarrow{h} _{i,k}})
\end{equation}
where $||$ denotes vector concatenation.

As introduced in Section \ref{intro}, each word in the word list $\boldsymbol{WL_{i}} = \{wl_{i,1},wl_{i,2},...,wl_{i,K}\}$ of the $TSC_{i}$ does not contribute equally to determining whether $TSC_{i}$ is a spoiler or not.  For instance, words such character's name and verb related to survival or failure are highly likely to trigger spoilers, while common conjunctions, definite articles or emotional adjectives do not have a great influence on spoiler judgments. Therefore, we equip word-level encoder with attention mechanism to select words that can accurately represent the sentence meaning. 

Given the Bi-LSTM state sequence $[\boldsymbol{h_{i,1},h_{i,2}, ...,h_{i,K}}]$, our attentive sequence encoder calculates a sequence of attention values $[\alpha_{i,1}, \alpha_{i,2}, ..., \alpha_{i,K}]$ by:
\begin{equation}
\alpha_{i,k} = \frac{exp(tanh(\boldsymbol{W_{s}\cdot h_{i,k})^{T}\cdot u_{s}})}{\sum_{t}exp(tanh(\boldsymbol{W_{s}\cdot h_{i,t})^{T}\cdot u_{s}})}
\end{equation}
where $\boldsymbol{W_{s}}$ is a weight matrix, and $\boldsymbol{u_{s}}$ is the context vector to distinguish informative words from non-informative ones. 

Finally, we obtain the target sentence vector $\boldsymbol{Tseq_{i}}$ of  $TSC_{i}$ by:
\begin{equation}
\boldsymbol{Tseq_{i}} = \sum_{k=1}^{K}\alpha_{i,k}\cdot \boldsymbol{h_{i,k}}
\end{equation}

According to the interactive property of TSCs, we consider the $R$ former neighbors $\boldsymbol{F} = \{F_{1},F_{2}, ... ,F_{R}\}$, and $P$ keyframes $\boldsymbol{KEY}=\{KEY_{1}, KEY_{2},...,KEY_{P}\}$ of of $TSC_{i}$. According to word-level attentive encoder, we obtain the sentence vector of former neighbors $\boldsymbol{NSEQ} =\{ \boldsymbol{Nseq_{1}}, \boldsymbol{Nseq_{2}}, ... $ $,\boldsymbol{Nseq_{R}}\}$ and keyframes $\boldsymbol{KSEQ} = \{\boldsymbol{Kseq_{1}}, \boldsymbol{Kseq_{2}}, ... , \boldsymbol{Kseq_{P}}\}$ based on certain TSCs which represent the semantic of keyframes, and take them as the input to the following SBN.

\subsection{Similarity-Based Network}
\label{4.2}
In this section, we introduce the SBN to obtain the likelihood of spoiler based on the semantic similarity between the target and its neighbors and the target and the keyframes.

Fig. \ref{fig:seq} shows our integrated SBN. As mentioned in Section \ref{4.2}, after the process of word-level attentive encoder, we obtain $R$ consecutive sequence vectors of former neighbors $\boldsymbol{NSEQ}$=$\{\boldsymbol{Nseq_{1}}$, $\boldsymbol{Nseq_{2}}$, ...  ,$\boldsymbol{Nseq_{R}\}}$ with their corresponding timestamps $t_{F_{1}} \leq t_{F_{2}} \leq ... \leq t_{F_{R}}$. For keyframes, we also use the word-level attentive encoder to get the semantic vector of each TSC, average the semantic vectors of all the TSCs in one keyframe and get the semantic sequence of $P$ keyframes $\boldsymbol{KSEQ= \{Kseq_{1},  ... , Kseq_{P}\}}$. 

We first compute neighbor similarity $Nsim_{r}$ between the target $TSC_{i}$ and each of its former neighbors $F_{r}$ by 
\begin{equation}
Nsim_{r} = sim(\boldsymbol{Nseq_r, Tseq_i})
\end{equation}
where $sim(x,y) = \boldsymbol{\frac{x \cdot y}{|{x}| \cdot |{y}|}}$.

Then, we take the weighted average similarity between the target $TSC_{i}$ and its former neighbor by 
\begin{equation}
G_{Nsim_{i}} = \sum_{r=1}^{R}(Nsim_{r}\cdot decay(t_{F_{r}}, t_{i}))
\end{equation}
where 
\begin{equation}
decay(t_{F_{r}}, t_{i}) = \frac{exp(-\beta(t_{F_{r}}-t_{i}))}{\sum_{k=1}^{R}exp(-\beta(t_{F_{k}}-t_{i}))}
\label{decay}
\end{equation}
is the time decay function expressing that the interactive property between two TSCs decreases with the increase of time interval. If a former neighbor has a long time interval with the target TSC, it has a small effect on the overall neighbor similarity. Time decay rate $\beta$ is a hyperparameter that will be discussed in Section \ref{5}.

Meanwhile, we compute keyframe similarity $Ksim_{p}$ between target $TSC_{i}$ and each keyframe $KEY_{p}$. To be noticed, we do not introduce time decay here because we do not care the time interval between keyframe and the target comment. Also, as long as the target TSC is similar to any keyframe, it is likely to be a spoiler. Therefore, we take the maximum keyframe similarity of the target $TSC_{i}$ as the overall keyframe similarity $G_{Ksim_{i}}$.

\begin{equation}
Ksim_{p} = sim(\boldsymbol{Kseq_p, Tseq_i})
\end{equation}
\begin{equation}
G_{Ksim_{i}} = Max.\{Ksim_{1}, Ksim_{2}, ... ,Ksim_{P}\}
\end{equation}
where $Max.$ denotes the maximum operation.

In our work, we take the prediction of the likelihood of spoiler as a binary classification problem, where 1 means the target TSC is a spoiler, and 0 otherwise. Intuitively, if the semantic of target TSC is closer to each of its former neighbors, it is less likely to be a spoiler because its topic describes current video content. Moreover, if the semantics of the target TSC is closer to any keyframe, it has more possibilities to be a spoiler because it is more likely to talk about important plots rather than following the playback time of the video. Therefore, we compute the difference between the overall neighbor similarity and overall keyframe similarity as a result. We determine the prediction result as
\begin{equation}
\hat{y}_{i}= sigmoid( G_{Ksim_{i}} - G_{Nsim_{i}} )
\end{equation}
where $sigmoid(x)=\frac{1}{1+e^{-x}}$.

Finally, we use binary cross-entropy as our loss function:
\begin{equation}
L = y_{i}\cdot ln \hat{y}_{i} + (1-y_{i}) \cdot ln(1-\hat{y_{i}})
\end{equation}
where $y_{i}$ is the ground truth that would be 1 if the TSC is a spoiler, and 0 otherwise.

\subsection{Sentence-Level Interactive Variance Attention}
\label{4.3}
In section \ref{4.2}, we have calculated the overall neighbor similarity based on the weighted average similarity between the target TSC and its former neighbors. However, the TSC data is high-noise. If the noise is contained in former neighbors, it makes trouble to generate accurate overall neighbor similarity. Therefore, we implement the sentence-level IVA mechanism to effectively eliminate the impact of noise. IVA mechanism aims to detect noise though similarity variance and assign a lower weight to the noise.

\begin{figure}[htbp]
  \centering
  \includegraphics[width=1\linewidth]{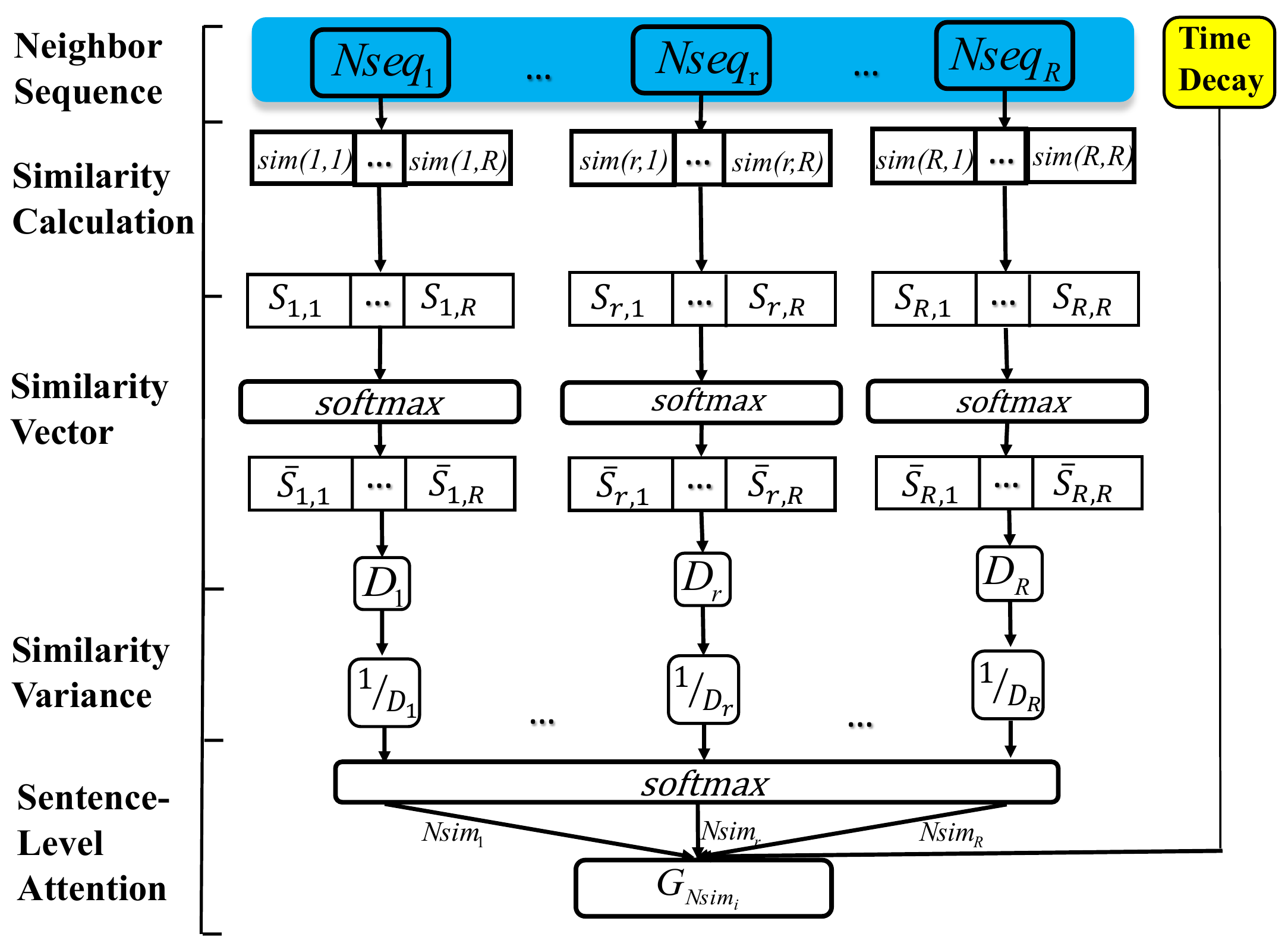}
  \caption{Sentence-Level Interactive Variance Attention (IVA) Framework.}\label{fig:atta}
\end{figure}

The IVA mechanism is formalized into a deep learning framework shown in Fig. \ref{fig:atta}. For each target $TSC_{i}$, we feed its neighbor sequence $\{Nseq_{1},$ $ Nseq_{2}, ... ,Nseq_{R}\}$ as the input of the framework.

Since noise has weak relevances with their surroundings \cite{Yang17}, we compute the similarity of any two TSCs in the sequence. For each $Nseq_{r}$, we compute its similarity with each TSC in the whole input sequence as the similarity vector $ \boldsymbol{S_{r}}=\{S_{r,1},S_{r,2},...,S_{r,R} \}$ by 
\begin{equation}
S_{i,j}=sim(\boldsymbol{Nseq_{i}},\boldsymbol{Nseq_{j}})
\end{equation}
and normalize it by:
\begin{equation}
\overline S_{i,j}= softmax(S_{i,j})
\end{equation}
where $softmax(x_{i})=\frac{exp(x_{i})}{\sum_{j}exp(x_{j})}$.

The noise has weak relevance to its surroundings and has a high similarity to itself, so the similarity distribution of noise is very concentrated and have relatively high similarity variance. On the contrary, non-noise TSCs have relatively placid similarity distribution and low similarity variance. Based on this point, we can distinguish noise comments from non-noise comments through their difference in similarity variance.

For $Nseq_{r}$, we compute the similarity variance $D_{r}$ of its normalized similarity vector $  \boldsymbol{  \overline S_{r}}  = \{\overline S_{r,1}, \overline S_{r,2}, ..., \overline S_{r,R} \}$ as:
\begin{equation}
D_{r} = \frac{1}{R}\sum_{i=1}^{R}(\overline S_{r,i} - Ave.( \boldsymbol{\overline S_{r}} ))^{2}
\end{equation}
where $Ave.( \boldsymbol{\overline S_{r}} )$ is the average of normalized similarity vector $ \boldsymbol{ \overline S_{r}}$.

We normalize $\frac{1}{D_{r}}$ through softmax function:
\begin{equation}
\overline D_{r} = softmax(\frac{1}{D_r})
\end{equation}
as the attention scores because higher variance means weaker relevance with surrounding comments. 
 
Finally, we compute the overall neighbor similarity $G_{Nsim_{r}}$ by the weighted sum of neighbor similarities:
\begin{equation}
G_{Nsim_{i}} = \sum_{r=1}^{R} \overline D_{r} \cdot Nsim_{r} \cdot decay(t_{F_{r}}, t_{i})
\end{equation}

Through sentence-level IVA, we make use of the interactive and real-time properties of TSC data to reduce the impact of noise effectively, and acquire the overall neighbor similarity more accurately.

\section{Experiment}
\label{5}
In this section, we first introduce dataset we used in this work in Section \ref{5.0} and the dataset process and evaluation metric in Section \ref{5.1}. Then, we evaluate the effectiveness of our proposed method by comparing with 4 baseline methods in Section \ref{5.2}. Finally, we visualize the weight of the attention layer to validate the performance of the attention mechanism both in word and sentence level in Section \ref{5.3}.

\subsection{Data Preparation}
\label{5.0}
The TSC data used in this paper is crawled from Chinese video website Youku \footnote{http://www.youku.com/} and Bilibili \footnote{http://www.bilibili.com}. We choose TV-series, movies, and sports as the categories of our data set, since the video plots of these three categories are very likely to be spoiled. These spoilers are relatively simple to be detected and labeled though certain spoiler keywords related to the plots. We gather the data from November 2017 to March 2018, where each TSC has been labeled as a spoiler or not by man-selected keyword and manual inspection.

To label the data, we first summarize the possible spoiler keywords specific to each video's tags and filter out the TSCs that most likely to be spoilers based on keywords. Then three human evaluators check the high-likely spoilers and assign labels to real spoilers according to whether the TSC is related to the latter plots of the video. We treat a certain TSC as a spoiler if two or more evaluators regard it related to the crucial plot. To better understand the insights of this dataset, we conduct preliminary statistic analyses listed as Table \ref{tab:spoilerdata}. The average density of TSCs means the average number of TSCs appear per second.

\begin{table}[htbp]
\centering
  \caption{Statistics of spoilers in the time-synchronized comment (TSC) dataset}
  \label{tab:spoilerdata}
  \begin{tabular}{llll}
    \toprule
    &TV-series&Movies&Sports\\
    \midrule
     Number of videos&32&480&501\\
    Number of TSCs&337568&591254&588036\\
    Ave. TSCs per video&10549.0&1231.8&1173.7\\
    Number of spoilers&102677&99868&98204\\
    Proportion of spoiler&0.3044&0.1689&0.1670\\
    Ave. Spoilers per video&3208.7&208.1&196.0\\
    Ave. Density of TSCs&4.14&0.9223&0.8687\\
  \bottomrule
\end{tabular}
\end{table}

\subsection{Dataset Preprocess and Evaluation Metric}
\label{5.1}
The raw TSC text is full of noise, so we manually remove meaningless TSCs and establish a set of mapping rules for network slang, which will be substituted by their real meaning in the text. For instance, 233... (2 followed by several 3) means laughter, 666... (several 6) means playing games very well, and ``front high energy'' means terrible plots will happen next in the video. After that, we segment the words and remove the anomaly symbol in TSCs by an open-source Chinese-language processing toolbox Jieba.

Also, timestamps are recorded when users send their TSCs at the video's playback time. To avoid a long time interval between consecutive TSCs and take better advantage of the inner relationships among TSCs, we discard the videos with fewer than 300 TSCs or its density of TSCs is lower than 0.1 comment per second. The final statistics of the data set have been introduced in Table \ref{tab:spoilerdata}.  We randomly select 70\% of the dataset as the training set, 20\% as the test set, and 10\% as the validation set.

In our model, we set $R = 5$ and $P = 3$, i.e, in each video, we sample 5 former neighbors of each target TSC and select 3 keyframes. Moreover, the time decay rate $\beta$ is hyper-parameter need to be decided. In the training progress, we adjust the value of $\beta$ by the validation set. The initial learning rate of Adam is 0.001, and the word vector dimension is 128. We get the best results when $\beta=0.15$, and will discuss them in details in Section \ref{5.2}.

The prediction of the likelihood of spoilers is a binary classification problem and the positive sample ratio (spoiler ratio) only accounts for 19.83\% of the total samples. Therefore, only considering the prediction performance of positive samples can better test the performance of the models. Thus, we use Precision, Recall and F1-score instead of Accuracy-score to measure the performance.

\subsection{Comparison of Baseline Methods}
\begin{table}
  \caption{F1-score, Precision, and Recall of each method}
  \label{tab:evalution}
  \begin{tabular}{llllllll}
    \multicolumn{4}{c}{\textbf{(a) TV-series}}\\
    \toprule
   Baselines & Precision & Recall & F1-score\\
        \midrule
    KM & 0.443 & \textbf{0.892} & 0.577\\
    LDA & 0.563 & 0.656 & 0.606 \\
    LI-NPP & 0.677 & 0.713 & 0.695\\
    DN-GAA & 0.730 & 0.798 & 0.747\\
    \hline
    SBN & 0.782 & 0.791 & 0.779 \\
    SBN-WT & 0.762& 0.743&  0.754\\
    SBN-IVA & \textbf{0.843} & 0.856 & \textbf{0.850} \\
      \bottomrule
  \\
  \multicolumn{4}{c}{\textbf{(b) Movies}}\\
    \toprule
    Baselines& Precision & Recall & F1-score\\
        \midrule
    KM & 0.398 & \textbf{0.912} & 0.545\\
    LDA & 0.497 & 0.604 & 0.562 \\
    LI-NPP & 0.654 & 0.727 & 0.688\\
    DN-GAA & 0.714 & 0.792 & 0.751\\
        \hline
    SBN & 0.722 & 0.833 & 0.785 \\
    SBN-WT &0.708 & 0.776 & 0.732 \\
    SBN-IVA & \textbf{0.753} & 0.856 & \textbf{0.811} \\
      \bottomrule
  \\
  \multicolumn{4}{c}{\textbf{(c) Sports}}\\
    \toprule
    Baselines& Precision & Recall & F1-score\\
        \midrule
    KM & 0.373 & 0.797 & 0.509\\
    LDA & 0.525 & 0.589 & 0.555 \\
    LI-NPP & 0.668 & 0.778 & 0.719\\
    DN-GAA & 0.718 & 0.758 & 0.738\\
        \hline
    SBN & 0.782 & 0.809 & 0.789 \\
    SBN-WT & 0.722 & 0.749 & 0.741\\
    SBN-IVA & \textbf{0.810} &\textbf{0.841} & \textbf{0.825} \\
      \bottomrule
\end{tabular}
\end{table}

\begin{table*}[!htbp]
\centering
\caption{F1-score, Precision, and Recall with different IVA parameters}
\label{tab:param}
\begin{tabular}{c|ccc|ccc|ccc|ccc}
\toprule 
Parameters&\multicolumn{3}{|c|}{R=1} &\multicolumn{3}{|c|}{R=2} & \multicolumn{3}{|c}{R=3} & \multicolumn{3}{|c}{R=4} \\
\hline
Metrics & Precision & Recall & F1-score & Precision & Recall & F1-score  & Precision & Recall & F1-score & Precision & Recall & F1-score\\
\hline
P=1 & 0.489&0.472&0.480  & 0.525&0.566&0.525  & 0.537&0.589&0.540 & 0.545&0.607&0.555\\
P=2 & 0.583&0.563&0.561 & 0.638&0.662&0.649  & 0.642&0.682&0.652  & 0.628&0.695&0.673\\
P=3   & 0.688&0.733&0.712 & 0.759&0.762&0.740  & 0.817&0.797&0.807 & 0.805& 0.811 & 0.807\\
P=4  & 0.730 & 0.762 & 0.744 & 0.770&0.824 & 0.797& 0.822 & 0.833 & 0.819 &0.829& 0.828 & 0.814\\
P=5  & 0.777 & 0.781 & 0.769  & 0.817 & 0.845& 0.841  & 0.825&0.860&0.846 & 0.819& 0.869 & 0.850\\
P=6   & 0.779 & 0.808 &0.773& 0.820 & 0.853 & 0.839 & 0.830 & 0.854 & 0.845& 0.827& 0.855 & 0.844\\
\bottomrule
\end{tabular}
\end{table*}

\label{5.2}
To evaluate the performance of our proposed model, we compare our models with the following methods as baselines:
\begin{itemize}

\item \textbf{KM:} Keyword-matching method is the simplest method, which filters out spoilers according to the actor name in a drama or the match score at a sporting game \cite{Golbeck12,Nakamura07}. We set the prediction result as 1, if a TSC matches spoiler keywords, and 0 otherwise.

\item \textbf{LDA:} This is a machine-learning method based on Latent Dirichlet Allocation (LDA) used by Guo \emph{et al.} \cite{Guo10} in their work to filter out spoiler using predictive perplexity. We use perplexity to predict the appropriate number of topics, and set it as 20.

\item \textbf{LI-NPP:} This method introduces location information (LI) and neighborhood plot probability (NPP) to help identify spoilers with contextual information through SVM \cite{Yoshinori16}. We set a linear kernel and complexity constant $C = 1$ in the SVM.

\item \textbf{DN-GAA:} This is a Deep Neural spoiler detection model (DN) that uses a Genre-Aware Attention mechanism (GAA) \cite{chang2018deep}.

\item \textbf{SBN:} Our Similarity-Based Network that proposed in Section \ref{4.2}.

\item \textbf{SBN-WT:} Our Similarity-Based Network Without Timestamps. In this model, we don't consider the timestamp and set the decay function (Eq.(\ref{decay})) equal to 1, which is used to evaluate the contribution of timestamp and decay function.

\item \textbf{SBN-IVA:} Our Similarity-Based Network works with Interactive Variance Attention mechanism that proposed in Section \ref{4.3}.

\end{itemize}

The baseline methods may reach their best performance with different latent factors. Therefore, we tested various parameters in the range of their work and chose the best prediction performance as final results for each baseline method.

The experiments of all the models are repeated for 10 times, and we use the average values as the final results. The results of F1-score, Precision, and Recall in the categories of TV-series, movies and sports are shown in table \ref{tab:evalution}.

From Table \ref{tab:evalution}, we can conclude that SBN-IVA achieves the best performance on Precision, Recall and F1-Score in the categories of TV-series, movies and sports. Compared to the-state-of-art DN-GAA method proposed by Chang \emph{et al} \cite{chang2018deep}, SBN-IVA enhance the performance 13.8\%, 7.99\%, 11.8\% (average 11.2\%) upon F1-Score. In other baselines, keyword-matching (KM) method achieves high Recall and low Precision, since it treats many non-spoiler TSCs as spoilers. Latent Dirichlet Allocation (LDA) based method does not perform well because the LDA model is not suitable to process short texts like TSC data. LI-NPP is an SVM-based method and has better performance than the unsupervised learning methods. DN-GAA is the state-of-the-art method and has the highest performance among the baselines. The results show that the neural network has powerful feature extraction capabilities when there is sufficient data. 

Although LI-NPP and DN-GAA have achieved better results than unsupervised learning methods, they do not consider the specific properties of TSC. Our SBN framework takes the interactive and real-time properties of TSCs into considering, and performs better than LI-NPP and DN-GAA. To further evaluate the effect of the real-time property of TSCs, we remove the decay function in SBN (set the decay function Eq.(\ref{decay}) equal to 1), and obtain SBN-WT. Results show that compared with SBN-WT, SBN enhance the performance 3.31\%, 7.24\%, 6.47\% upon F1-Score. This indicates that it is necessary to provide a low weight to those TSCs that has a long time interval from the target TSC because they maybe discuss different topics. Moreover, to evaluate the contribution of IVA mechanism, we compare the SBN with SBN-IVA. Results show that SBN-IVA enhances the performance  9.11\%, 3.31\%, 4.56\% upon F1-Score than SBN, which indicates that the IVA mechanism can reduce the impact of noise efficiently. Our SBN-IVA achieves the best Precision, Recall, and F1-Score, because our model effectively makes use of the interactive and real-time properties of TSCs and reduce the impact of noise.

\begin{figure}[htbp]
  \centering
  \includegraphics[width=1\linewidth]{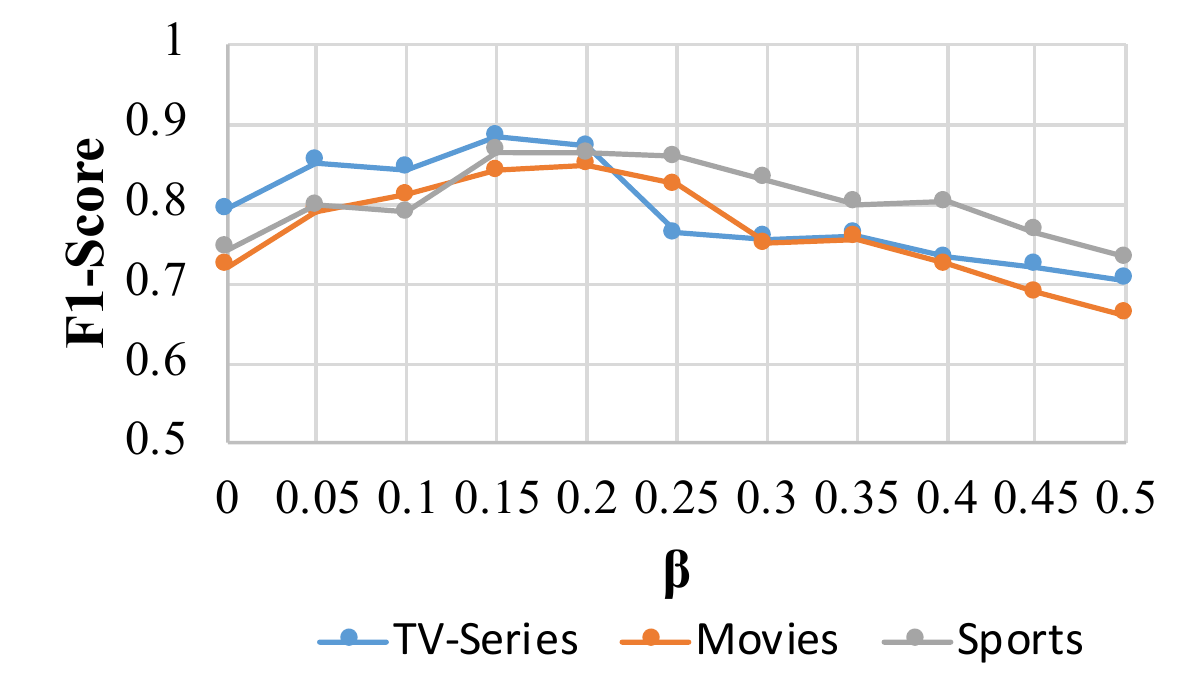}
  \caption{The influence of time decay rate $\beta$.}\label{fig:f1}
\end{figure}

Then, we discuss the influence of time decay rate $\beta$ on the experimental results. $\beta=0$ means we do not consider the time decay. When time decay rate $\beta$ grows, the interactive property of TSCs becomes weaker with the time interval. We change the value of $\beta$ from 0 to 0.5 with the step size of 0.05 and calculate the F1-score in the categories of TV-series, movies, and sports. As shown in Figure \ref{fig:f1}, the decay rate $\beta$ reaches its best performance when $\beta=0.15$.

Finally, we change the number of former neighbors $R$ and keyframe $P$ to see their influence on the experiment result in the validation set. The results of F1-score, Precision, and Recall with different IVA parameters are shown in Table \ref{tab:param}. We can find the F1-score increase with the number of former neighbors $R$ and keyframe $P$. This result proves that our application of the interactive and real-time properties of TSCs is correct. When we increase former neighbors, TSCs establish semantics association with more surrounding neighbors. When we increase keyframes, TSCs can be compared with more important plots in the video. According to table \ref{tab:param}, we choose $P=5$ and $R=3$, and do not continue increasing $P$ and $R$ because this leads to a decline in time efficiency and there is no significant growth in Precision, Recall, and F1-score.

\subsection{Visualization of Attention}
\label{5.3}
\begin{figure}[htbp]
  \centering
  \includegraphics[width=1\linewidth]{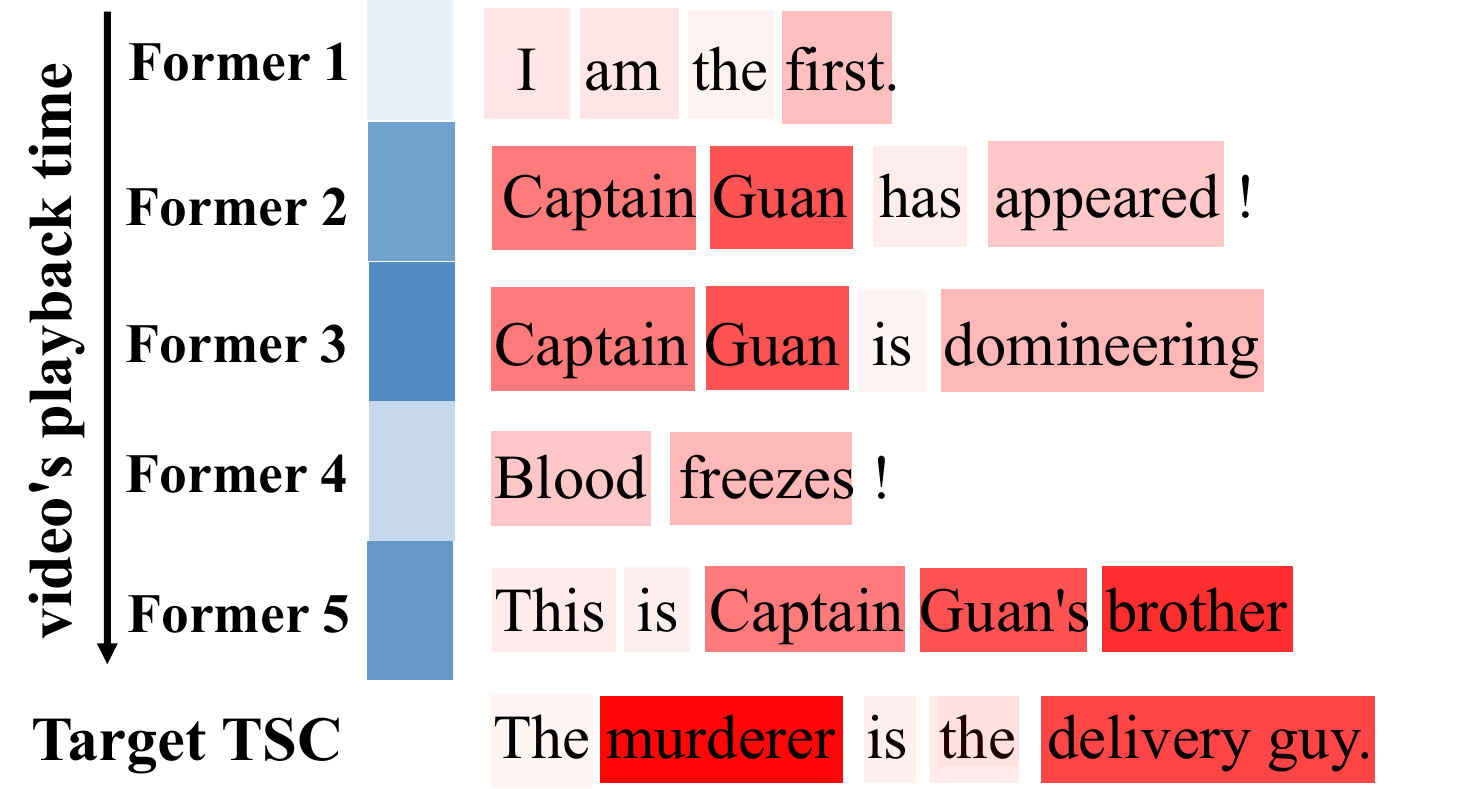}
  \caption{Visualization of attention in word and sentence level. }\label{fig:visual}
\end{figure}

To validate the performance of our attention mechanism both in word and sentence level, we visualize the weight of the attention layer in Fig. \ref{fig:visual}. We take a group of consecutive TSCs in a criminal investigation TV Series as an example. The scene of the example is shown in Fig. \ref{intro}.

The TSC in the last line is the target TSC, and the others are its former neighbors in order of timestamps. The blue color denotes the weight of sentences in sentence-level IVA mechanism. The darker the color is, the higher the IVA score the TSC has.  Meanwhile, the red color denotes the weight of words in the word-level attentive encoder. The darker the color is, the more likely the word to be a keyword to define a spoiler.

Fig. \ref{fig:visual} shows the effectiveness of our attention model. We can see that these TSCs mainly discuss something about Captain Guan (the protagonist in the TV-series). The weight of sentence-level IVA is calculated by the product of semantic similarity and time decay. The Former 2, Former 3, and Former 5 have high IVA scores, which means they are related to the main topic. In contrast, Former 1 and Former 4 mainly express the users' emotion deviated to the central topic, therefore they should be classified as noise and have a small weight.

In the word-level, Fig \ref{fig:visual} shows that our attentive word encoder can select the words carrying strong spoiler tendencies such as ``Captain'', ``Guan'', ``brother'', ``murderer'', and ``delivery guy''. Other unimportant words such as``am'', ``is'', ``this'', ``has'', are discarded. That is, the word attentive encoder selects spoiler keyword when calculating the sequence vector through the weighted sum.

\section{Conclusion and future work}
In this paper, we propose a novel similarity-based spoiler detection model with the dataset of TSCs. To extract spoiler keyword accurately based on the short-text, interactive, real-time and noise properties of TSCs, we introduce attention mechanism both in word encoder and sentence invariance level in our model. The Word-Level Attentive Encoder filters the spoiler keyword out, and Interactive Variance Attention (IVA) calculate the similarity invariance of comments and assign a low weight to noise comments. In this way, we integrate the short-text, interactive and real-time properties of TSCs in the deep framework, and effectively weaken the impact of noise comments. Then the likelihood of TSCs to be spoilers is predicted accurately through the difference of neighbor similarity and keyframe similarity. Extensive experiments on real-world dataset proved that our model with IVA outperformed the existing spoiler detection method in F1-score, Precision, and Recall.

Our pioneer work in this paper points out the promising future directions in online spoiler detection, and this is the first step towards our goal in online spoiler detection. In the future, we will seek to explore the following directions: (1) We will take other information about the online video, such as visual and audio features to improve the performance of the model. (2) In most cases, the spoilers are sparse. However, some extreme users will continue to post the spoilers to occupy the screen. In this extreme case, our model cannot solve it well (since all the neighbors are the same and post by extreme users). Therefore, we will incorporate user ID into the model to identify extreme spoilers in the next step. (3) We will design an online spoiler detection model to detect video spoilers in real time.

%

%
\begin{acks}
This work is supported by Chinese National Research Fund (NSFC) Key Project No. 61532013 and No. 61872239; 0007/2018/A1, 0060/2019/A1, DCT-MoST Joint-project No. 025/2015/AMJ of Science and Technology Development Fund, Macao S.A.R (FDCT), China, and University of Macau Grant Nos: MYRG2018-00237-RTO, CPG2019-00004-FST and SRG2018-00111-FST.
\end{acks}

%
\bibliographystyle{ACM-Reference-Format}
\bibliography{CIKM-sigconf}


\begin{thebibliography}{37}


\ifx \showCODEN    \undefined \def \showCODEN     #1{\unskip}     \fi
\ifx \showDOI      \undefined \def \showDOI       #1{#1}\fi
\ifx \showISBNx    \undefined \def \showISBNx     #1{\unskip}     \fi
\ifx \showISBNxiii \undefined \def \showISBNxiii  #1{\unskip}     \fi
\ifx \showISSN     \undefined \def \showISSN      #1{\unskip}     \fi
\ifx \showLCCN     \undefined \def \showLCCN      #1{\unskip}     \fi
\ifx \shownote     \undefined \def \shownote      #1{#1}          \fi
\ifx \showarticletitle \undefined \def \showarticletitle #1{#1}   \fi
\ifx \showURL      \undefined \def \showURL       {\relax}        \fi
\providecommand\bibfield[2]{#2}
\providecommand\bibinfo[2]{#2}
\providecommand\natexlab[1]{#1}
\providecommand\showeprint[2][]{arXiv:#2}

\bibitem[\protect\citeauthoryear{Bahdanau, Cho, and Bengio}{Bahdanau
  et~al\mbox{.}}{2014}]%
        {Bahdanau14}
\bibfield{author}{\bibinfo{person}{Dzmitry Bahdanau},
  \bibinfo{person}{Kyunghyun Cho}, {and} \bibinfo{person}{Yoshua Bengio}.}
  \bibinfo{year}{2014}\natexlab{}.
\newblock \showarticletitle{Neural machine translation by jointly learning to
  align and translate}. In \bibinfo{booktitle}{\emph{arXiv preprint
  arXiv:1409.0473}}.
\newblock


\bibitem[\protect\citeauthoryear{Bojar, Chatterjee, Federmann, Graham, Haddow,
  Huang, Huck, Koehn, Liu, Logacheva, et~al\mbox{.}}{Bojar
  et~al\mbox{.}}{2017}]%
        {bojar2017findings}
\bibfield{author}{\bibinfo{person}{Ond{\v{r}}ej Bojar}, \bibinfo{person}{Rajen
  Chatterjee}, \bibinfo{person}{Christian Federmann}, \bibinfo{person}{Yvette
  Graham}, \bibinfo{person}{Barry Haddow}, \bibinfo{person}{Shujian Huang},
  \bibinfo{person}{Matthias Huck}, \bibinfo{person}{Philipp Koehn},
  \bibinfo{person}{Qun Liu}, \bibinfo{person}{Varvara Logacheva},
  {et~al\mbox{.}}} \bibinfo{year}{2017}\natexlab{}.
\newblock \showarticletitle{Findings of the 2017 conference on machine
  translation (wmt17)}. In \bibinfo{booktitle}{\emph{Proceedings of the Second
  Conference on Machine Translation}}. \bibinfo{pages}{169--214}.
\newblock


\bibitem[\protect\citeauthoryear{Chang, Kim, Kim, Kim, and Kang}{Chang
  et~al\mbox{.}}{2018}]%
        {chang2018deep}
\bibfield{author}{\bibinfo{person}{Buru Chang}, \bibinfo{person}{Hyunjae Kim},
  \bibinfo{person}{Raehyun Kim}, \bibinfo{person}{Deahan Kim}, {and}
  \bibinfo{person}{Jaewoo Kang}.} \bibinfo{year}{2018}\natexlab{}.
\newblock \showarticletitle{A Deep Neural Spoiler Detection Model Using a
  Genre-Aware Attention Mechanism}. In \bibinfo{booktitle}{\emph{Pacific-Asia
  Conference on Knowledge Discovery and Data Mining}}. Springer,
  \bibinfo{pages}{183--195}.
\newblock


\bibitem[\protect\citeauthoryear{Chen, Chen, Ma, Yao, Liu, Luo, and Zhang}{Chen
  et~al\mbox{.}}{2018}]%
        {chen2018fine}
\bibfield{author}{\bibinfo{person}{Xinpeng Chen}, \bibinfo{person}{Jingyuan
  Chen}, \bibinfo{person}{Lin Ma}, \bibinfo{person}{Jian Yao},
  \bibinfo{person}{Wei Liu}, \bibinfo{person}{Jiebo Luo}, {and}
  \bibinfo{person}{Tong Zhang}.} \bibinfo{year}{2018}\natexlab{}.
\newblock \showarticletitle{Fine-grained video attractiveness prediction using
  multimodal deep learning on a large real-world dataset}. In
  \bibinfo{booktitle}{\emph{Companion of the The Web Conference 2018 on The Web
  Conference 2018}}. International World Wide Web Conferences Steering
  Committee, \bibinfo{pages}{671--678}.
\newblock


\bibitem[\protect\citeauthoryear{Chen, Zhang, Ai, Xu, Yan, and Qin}{Chen
  et~al\mbox{.}}{2017}]%
        {chen2017personalized}
\bibfield{author}{\bibinfo{person}{Xu Chen}, \bibinfo{person}{Yongfeng Zhang},
  \bibinfo{person}{Qingyao Ai}, \bibinfo{person}{Hongteng Xu},
  \bibinfo{person}{Junchi Yan}, {and} \bibinfo{person}{Zheng Qin}.}
  \bibinfo{year}{2017}\natexlab{}.
\newblock \showarticletitle{Personalized key frame recommendation}. In
  \bibinfo{booktitle}{\emph{Proceedings of the 40th International ACM SIGIR
  Conference on Research and Development in Information Retrieval}}. ACM,
  \bibinfo{pages}{315--324}.
\newblock


\bibitem[\protect\citeauthoryear{Chen}{Chen}{2017}]%
        {Chen17}
\bibfield{author}{\bibinfo{person}{Yixin Chen}.}
  \bibinfo{year}{2017}\natexlab{}.
\newblock \showarticletitle{Automatic Annotation of Online Multimedia Data with
  Similarity Relations}.
\newblock \bibinfo{journal}{\emph{McGill University}} (\bibinfo{year}{2017}).
\newblock


\bibitem[\protect\citeauthoryear{Devlin, Chang, Lee, and Toutanova}{Devlin
  et~al\mbox{.}}{2018}]%
        {devlin2018bert}
\bibfield{author}{\bibinfo{person}{Jacob Devlin}, \bibinfo{person}{Ming-Wei
  Chang}, \bibinfo{person}{Kenton Lee}, {and} \bibinfo{person}{Kristina
  Toutanova}.} \bibinfo{year}{2018}\natexlab{}.
\newblock \showarticletitle{Bert: Pre-training of deep bidirectional
  transformers for language understanding}.
\newblock \bibinfo{journal}{\emph{arXiv preprint arXiv:1810.04805}}
  (\bibinfo{year}{2018}).
\newblock


\bibitem[\protect\citeauthoryear{Golbeck}{Golbeck}{2012}]%
        {Golbeck12}
\bibfield{author}{\bibinfo{person}{Jennifer Golbeck}.}
  \bibinfo{year}{2012}\natexlab{}.
\newblock \showarticletitle{The twitter mute button: a web filtering
  challenge}. In \bibinfo{booktitle}{\emph{Proceedings of the SIGCHI Conference
  on Human Factors in Computing Systems}}. ACM, \bibinfo{pages}{2755--2758}.
\newblock


\bibitem[\protect\citeauthoryear{Guo and Ramakrishnan}{Guo and
  Ramakrishnan}{2010}]%
        {Guo10}
\bibfield{author}{\bibinfo{person}{Sheng Guo} {and} \bibinfo{person}{Naren
  Ramakrishnan}.} \bibinfo{year}{2010}\natexlab{}.
\newblock \showarticletitle{Finding the storyteller: automatic spoiler tagging
  using linguistic cues}. In \bibinfo{booktitle}{\emph{Proceedings of the 23rd
  International Conference on Computational Linguistics}}. ACL,
  \bibinfo{pages}{412--420}.
\newblock


\bibitem[\protect\citeauthoryear{He, Ge, Wu, Chen, and Tan}{He
  et~al\mbox{.}}{2016}]%
        {He16}
\bibfield{author}{\bibinfo{person}{Ming He}, \bibinfo{person}{Yong Ge},
  \bibinfo{person}{Le Wu}, \bibinfo{person}{Enhong Chen}, {and}
  \bibinfo{person}{Chang Tan}.} \bibinfo{year}{2016}\natexlab{}.
\newblock \showarticletitle{Predicting the popularity of danmu-enabled videos:
  A multi-factor view}. In \bibinfo{booktitle}{\emph{International Conference
  on Database Systems for Advanced Applications}}. Springer,
  \bibinfo{pages}{351--366}.
\newblock


\bibitem[\protect\citeauthoryear{Hijikata, Iwai, and Nishida}{Hijikata
  et~al\mbox{.}}{2016}]%
        {Yoshinori16}
\bibfield{author}{\bibinfo{person}{Yoshinori Hijikata},
  \bibinfo{person}{Hidenari Iwai}, {and} \bibinfo{person}{Shogo Nishida}.}
  \bibinfo{year}{2016}\natexlab{}.
\newblock \showarticletitle{Context-Based Plot Detection from Online Review
  Comments for Preventing Spoilers}. In \bibinfo{booktitle}{\emph{2016
  IEEE/WIC/ACM International Conference on Web Intelligence (WI)}}. IEEE,
  \bibinfo{pages}{57--65}.
\newblock


\bibitem[\protect\citeauthoryear{Hochreiter and Schmidhuber}{Hochreiter and
  Schmidhuber}{1997}]%
        {Hochreiter97}
\bibfield{author}{\bibinfo{person}{Sepp Hochreiter} {and}
  \bibinfo{person}{J{\"u}rgen Schmidhuber}.} \bibinfo{year}{1997}\natexlab{}.
\newblock \showarticletitle{Long short-term memory}.
\newblock \bibinfo{journal}{\emph{Neural computation}} \bibinfo{volume}{9},
  \bibinfo{number}{8} (\bibinfo{year}{1997}), \bibinfo{pages}{1735--1780}.
\newblock


\bibitem[\protect\citeauthoryear{Iwai, Hijikata, Ikeda, and Nishida}{Iwai
  et~al\mbox{.}}{2014}]%
        {Iwai14}
\bibfield{author}{\bibinfo{person}{Hidenari Iwai}, \bibinfo{person}{Yoshinori
  Hijikata}, \bibinfo{person}{Kaori Ikeda}, {and} \bibinfo{person}{Shogo
  Nishida}.} \bibinfo{year}{2014}\natexlab{}.
\newblock \showarticletitle{Sentence-based plot classification for online
  review comments}. In \bibinfo{booktitle}{\emph{2014 IEEE/WIC/ACM
  International Joint Conferences on Web Intelligence (WI) and Intelligent
  Agent Technologies (IAT)}}, Vol.~\bibinfo{volume}{1}. IEEE,
  \bibinfo{pages}{245--253}.
\newblock


\bibitem[\protect\citeauthoryear{Jeon, Kim, and Yu}{Jeon et~al\mbox{.}}{2013}]%
        {Jeon13}
\bibfield{author}{\bibinfo{person}{Sungho Jeon}, \bibinfo{person}{Sungchul
  Kim}, {and} \bibinfo{person}{Hwanjo Yu}.} \bibinfo{year}{2013}\natexlab{}.
\newblock \showarticletitle{Don’t be spoiled by your friends: spoiler
  detection in TV program tweets}. In \bibinfo{booktitle}{\emph{Seventh
  International AAAI Conference on Weblogs and Social Media}}.
\newblock


\bibitem[\protect\citeauthoryear{Kumar, Irsoy, Ondruska, Iyyer, Bradbury,
  Gulrajani, Zhong, Paulus, and Socher}{Kumar et~al\mbox{.}}{2016}]%
        {Kumar15}
\bibfield{author}{\bibinfo{person}{Ankit Kumar}, \bibinfo{person}{Ozan Irsoy},
  \bibinfo{person}{Peter Ondruska}, \bibinfo{person}{Mohit Iyyer},
  \bibinfo{person}{James Bradbury}, \bibinfo{person}{Ishaan Gulrajani},
  \bibinfo{person}{Victor Zhong}, \bibinfo{person}{Romain Paulus}, {and}
  \bibinfo{person}{Richard Socher}.} \bibinfo{year}{2016}\natexlab{}.
\newblock \showarticletitle{Ask me anything: Dynamic memory networks for
  natural language processing}. In \bibinfo{booktitle}{\emph{International
  Conference on Machine Learning}}. \bibinfo{pages}{1378--1387}.
\newblock


\bibitem[\protect\citeauthoryear{Li, Liao, Zhang, and Wang}{Li
  et~al\mbox{.}}{2016}]%
        {li2016event}
\bibfield{author}{\bibinfo{person}{Jiangfeng Li}, \bibinfo{person}{Zhenyu
  Liao}, \bibinfo{person}{Chenxi Zhang}, {and} \bibinfo{person}{Jing Wang}.}
  \bibinfo{year}{2016}\natexlab{}.
\newblock \showarticletitle{Event detection on online videos using crowdsourced
  time-sync comment}. In \bibinfo{booktitle}{\emph{2016 7th International
  Conference on Cloud Computing and Big Data (CCBD)}}. IEEE,
  \bibinfo{pages}{52--57}.
\newblock


\bibitem[\protect\citeauthoryear{Liao, Xian, Yang, Zhao, Zhang, and Li}{Liao
  et~al\mbox{.}}{2018}]%
        {liao2018tscset}
\bibfield{author}{\bibinfo{person}{Zhenyu Liao}, \bibinfo{person}{Yikun Xian},
  \bibinfo{person}{Xiao Yang}, \bibinfo{person}{Qinpei Zhao},
  \bibinfo{person}{Chenxi Zhang}, {and} \bibinfo{person}{Jiangfeng Li}.}
  \bibinfo{year}{2018}\natexlab{}.
\newblock \showarticletitle{TSCSet: A Crowdsourced Time-Sync Comment Dataset
  for Exploration of User Experience Improvement}. In
  \bibinfo{booktitle}{\emph{23rd International Conference on Intelligent User
  Interfaces}}. ACM, \bibinfo{pages}{641--652}.
\newblock


\bibitem[\protect\citeauthoryear{Luong, Pham, and Manning}{Luong
  et~al\mbox{.}}{2015}]%
        {Luong15}
\bibfield{author}{\bibinfo{person}{Minh-Thang Luong}, \bibinfo{person}{Hieu
  Pham}, {and} \bibinfo{person}{Christopher~D Manning}.}
  \bibinfo{year}{2015}\natexlab{}.
\newblock \showarticletitle{Effective approaches to attention-based neural
  machine translation}. In \bibinfo{booktitle}{\emph{arXiv preprint
  arXiv:1508.04025}}.
\newblock


\bibitem[\protect\citeauthoryear{Lv, Xu, Chen, Liu, and Zheng}{Lv
  et~al\mbox{.}}{[n. d.]}]%
        {Lv16}
\bibfield{author}{\bibinfo{person}{Guangyi Lv}, \bibinfo{person}{Tong Xu},
  \bibinfo{person}{Enhong Chen}, \bibinfo{person}{Qi Liu}, {and}
  \bibinfo{person}{Yi Zheng}.} \bibinfo{year}{[n. d.]}\natexlab{}.
\newblock \showarticletitle{Reading the Videos: Temporal Labeling for
  Crowdsourced Time-Sync Videos Based on Semantic Embedding}. In
  \bibinfo{booktitle}{\emph{Thirtieth AAAI Conference on Artificial
  Intelligence}}.
\newblock


\bibitem[\protect\citeauthoryear{Maeda, Hijikata, and Nakamura}{Maeda
  et~al\mbox{.}}{2016}]%
        {Kyosuke16}
\bibfield{author}{\bibinfo{person}{Kyosuke Maeda}, \bibinfo{person}{Yoshinori
  Hijikata}, {and} \bibinfo{person}{Satoshi Nakamura}.}
  \bibinfo{year}{2016}\natexlab{}.
\newblock \showarticletitle{A basic study on spoiler detection from review
  comments using story documents}. In \bibinfo{booktitle}{\emph{2016
  IEEE/WIC/ACM International Conference on Web Intelligence (WI)}}. IEEE,
  \bibinfo{pages}{572--577}.
\newblock


\bibitem[\protect\citeauthoryear{Mikolov, Chen, Corrado, and Dean}{Mikolov
  et~al\mbox{.}}{2013}]%
        {Mikolov2013Efficient}
\bibfield{author}{\bibinfo{person}{Tomas Mikolov}, \bibinfo{person}{Kai Chen},
  \bibinfo{person}{Greg Corrado}, {and} \bibinfo{person}{Jeffrey Dean}.}
  \bibinfo{year}{2013}\natexlab{}.
\newblock \showarticletitle{Efficient estimation of word representations in
  vector space}.
\newblock \bibinfo{journal}{\emph{arXiv preprint arXiv:1301.3781}}
  (\bibinfo{year}{2013}).
\newblock


\bibitem[\protect\citeauthoryear{Nakamura and Tanaka}{Nakamura and
  Tanaka}{2007}]%
        {Nakamura07}
\bibfield{author}{\bibinfo{person}{Satoshi Nakamura} {and}
  \bibinfo{person}{Katsumi Tanaka}.} \bibinfo{year}{2007}\natexlab{}.
\newblock \showarticletitle{Temporal filtering system to reduce the risk of
  spoiling a user's enjoyment}. In \bibinfo{booktitle}{\emph{Proceedings of the
  12th international conference on Intelligent user interfaces}}. ACM,
  \bibinfo{pages}{345--348}.
\newblock


\bibitem[\protect\citeauthoryear{Ping}{Ping}{2018}]%
        {ping2018video}
\bibfield{author}{\bibinfo{person}{Qing Ping}.}
  \bibinfo{year}{2018}\natexlab{}.
\newblock \showarticletitle{Video recommendation using crowdsourced time-sync
  comments}. In \bibinfo{booktitle}{\emph{Proceedings of the 12th ACM
  Conference on Recommender Systems}}. ACM, \bibinfo{pages}{568--572}.
\newblock


\bibitem[\protect\citeauthoryear{Ping and Chen}{Ping and Chen}{2017}]%
        {ping2017video}
\bibfield{author}{\bibinfo{person}{Qing Ping} {and} \bibinfo{person}{Chaomei
  Chen}.} \bibinfo{year}{2017}\natexlab{}.
\newblock \showarticletitle{Video Highlights Detection and Summarization with
  Lag-Calibration based on Concept-Emotion Mapping of Crowd-sourced Time-Sync
  Comments}.
\newblock \bibinfo{journal}{\emph{arXiv preprint arXiv:1708.02210}}
  (\bibinfo{year}{2017}).
\newblock


\bibitem[\protect\citeauthoryear{Shaw, Uszkoreit, and Vaswani}{Shaw
  et~al\mbox{.}}{2018}]%
        {shaw2018self}
\bibfield{author}{\bibinfo{person}{Peter Shaw}, \bibinfo{person}{Jakob
  Uszkoreit}, {and} \bibinfo{person}{Ashish Vaswani}.}
  \bibinfo{year}{2018}\natexlab{}.
\newblock \showarticletitle{Self-attention with relative position
  representations}.
\newblock \bibinfo{journal}{\emph{arXiv preprint arXiv:1803.02155}}
  (\bibinfo{year}{2018}).
\newblock


\bibitem[\protect\citeauthoryear{Sukhbaatar, Weston, Fergus,
  et~al\mbox{.}}{Sukhbaatar et~al\mbox{.}}{2015}]%
        {Sukhbaatar15}
\bibfield{author}{\bibinfo{person}{Sainbayar Sukhbaatar},
  \bibinfo{person}{Jason Weston}, \bibinfo{person}{Rob Fergus},
  {et~al\mbox{.}}} \bibinfo{year}{2015}\natexlab{}.
\newblock \showarticletitle{End-to-end memory networks}. In
  \bibinfo{booktitle}{\emph{Advances in neural information processing
  systems}}. \bibinfo{pages}{2440--2448}.
\newblock


\bibitem[\protect\citeauthoryear{Vaswani, Shazeer, Parmar, Uszkoreit, Jones,
  Gomez, Kaiser, and Polosukhin}{Vaswani et~al\mbox{.}}{2017}]%
        {Vaswani17}
\bibfield{author}{\bibinfo{person}{Ashish Vaswani}, \bibinfo{person}{Noam
  Shazeer}, \bibinfo{person}{Niki Parmar}, \bibinfo{person}{Jakob Uszkoreit},
  \bibinfo{person}{Llion Jones}, \bibinfo{person}{Aidan~N Gomez},
  \bibinfo{person}{{\L}ukasz Kaiser}, {and} \bibinfo{person}{Illia
  Polosukhin}.} \bibinfo{year}{2017}\natexlab{}.
\newblock \showarticletitle{Attention is all you need}. In
  \bibinfo{booktitle}{\emph{Advances in Neural Information Processing
  Systems}}. \bibinfo{pages}{5998--6008}.
\newblock


\bibitem[\protect\citeauthoryear{Vinyals, Kaiser, Koo, Petrov, Sutskever, and
  Hinton}{Vinyals et~al\mbox{.}}{2015}]%
        {Vinyals14}
\bibfield{author}{\bibinfo{person}{Oriol Vinyals}, \bibinfo{person}{{\L}ukasz
  Kaiser}, \bibinfo{person}{Terry Koo}, \bibinfo{person}{Slav Petrov},
  \bibinfo{person}{Ilya Sutskever}, {and} \bibinfo{person}{Geoffrey Hinton}.}
  \bibinfo{year}{2015}\natexlab{}.
\newblock \showarticletitle{Grammar as a foreign language}. In
  \bibinfo{booktitle}{\emph{Advances in neural information processing
  systems}}. \bibinfo{pages}{2773--2781}.
\newblock


\bibitem[\protect\citeauthoryear{Wu, Zhong, Tan, Horner, and Yang}{Wu
  et~al\mbox{.}}{2014}]%
        {Wu14}
\bibfield{author}{\bibinfo{person}{Bin Wu}, \bibinfo{person}{Erheng Zhong},
  \bibinfo{person}{Ben Tan}, \bibinfo{person}{Andrew Horner}, {and}
  \bibinfo{person}{Qiang Yang}.} \bibinfo{year}{2014}\natexlab{}.
\newblock \showarticletitle{Crowdsourced time-sync video tagging using temporal
  and personalized topic modeling}. In \bibinfo{booktitle}{\emph{Proceedings of
  the 20th ACM SIGKDD international conference on Knowledge discovery and data
  mining}}. ACM, \bibinfo{pages}{721--730}.
\newblock


\bibitem[\protect\citeauthoryear{Xian, Li, Zhang, and Liao}{Xian
  et~al\mbox{.}}{2015}]%
        {Xian15}
\bibfield{author}{\bibinfo{person}{Yikun Xian}, \bibinfo{person}{Jiangfeng Li},
  \bibinfo{person}{Chenxi Zhang}, {and} \bibinfo{person}{Zhenyu Liao}.}
  \bibinfo{year}{2015}\natexlab{}.
\newblock \showarticletitle{Video highlight shot extraction with time-sync
  comment}. In \bibinfo{booktitle}{\emph{Proceedings of the 7th International
  Workshop on Hot Topics in Planet-scale mObile computing and online Social
  neTworking}}. ACM, \bibinfo{pages}{31--36}.
\newblock


\bibitem[\protect\citeauthoryear{Xu and Zhang}{Xu and Zhang}{[n. d.]}]%
        {Xu17}
\bibfield{author}{\bibinfo{person}{Linli Xu} {and} \bibinfo{person}{Chao
  Zhang}.} \bibinfo{year}{[n. d.]}\natexlab{}.
\newblock \showarticletitle{Bridging Video Content and Comments: Synchronized
  Video Description with Temporal Summarization of Crowdsourced Time-Sync
  Comments}. In \bibinfo{booktitle}{\emph{Thirty-First AAAI Conference on
  Artificial Intelligence}}.
\newblock


\bibitem[\protect\citeauthoryear{Yang, Gao, Zhou, Jia, Zhang, and Luo}{Yang
  et~al\mbox{.}}{2019a}]%
        {Yang19}
\bibfield{author}{\bibinfo{person}{Wenmian Yang}, \bibinfo{person}{Wenyuan
  Gao}, \bibinfo{person}{Xiaojie Zhou}, \bibinfo{person}{Weijia Jia},
  \bibinfo{person}{Shaohua Zhang}, {and} \bibinfo{person}{Yutao Luo}.}
  \bibinfo{year}{2019}\natexlab{a}.
\newblock \showarticletitle{Herding Effect based Attention for Personalized
  Time-Sync Video Recommendatio}. In \bibinfo{booktitle}{\emph{2019 IEEE
  International Conference on Multimedia and Expo (ICME)}}. IEEE.
\newblock


\bibitem[\protect\citeauthoryear{Yang, Jia, Zhou, and Luo}{Yang
  et~al\mbox{.}}{2019b}]%
        {yang2019legal}
\bibfield{author}{\bibinfo{person}{Wenmian Yang}, \bibinfo{person}{Weijia Jia},
  \bibinfo{person}{XIaojie Zhou}, {and} \bibinfo{person}{Yutao Luo}.}
  \bibinfo{year}{2019}\natexlab{b}.
\newblock \showarticletitle{Legal Judgment Prediction via Multi-Perspective
  Bi-Feedback Network}. In \bibinfo{booktitle}{\emph{Proceedings of the 28th
  International Joint Conference on Artificial Intelligence (IJCAI)}}.
\newblock


\bibitem[\protect\citeauthoryear{Yang, Ruan, Gao, Wang, Ran, and Jia}{Yang
  et~al\mbox{.}}{2017}]%
        {Yang17}
\bibfield{author}{\bibinfo{person}{Wenmian Yang}, \bibinfo{person}{Na Ruan},
  \bibinfo{person}{Wenyuan Gao}, \bibinfo{person}{Kun Wang},
  \bibinfo{person}{Wensheng Ran}, {and} \bibinfo{person}{Weijia Jia}.}
  \bibinfo{year}{2017}\natexlab{}.
\newblock \showarticletitle{Crowdsourced time-sync video tagging using semantic
  association graph}. In \bibinfo{booktitle}{\emph{2017 IEEE International
  Conference on Multimedia and Expo (ICME)}}. IEEE, \bibinfo{pages}{547--552}.
\newblock


\bibitem[\protect\citeauthoryear{Yang, Wang, Ruan, Gao, Jia, Zhao, Liu, and
  Zhang}{Yang et~al\mbox{.}}{2019c}]%
        {yang2019time}
\bibfield{author}{\bibinfo{person}{Wenmian Yang}, \bibinfo{person}{Kun Wang},
  \bibinfo{person}{Na Ruan}, \bibinfo{person}{Wenyuan Gao},
  \bibinfo{person}{Weijia Jia}, \bibinfo{person}{Wei Zhao},
  \bibinfo{person}{Nan Liu}, {and} \bibinfo{person}{Yunyong Zhang}.}
  \bibinfo{year}{2019}\natexlab{c}.
\newblock \showarticletitle{Time-sync Video Tag Extraction Using Semantic
  Association Graph}.
\newblock \bibinfo{journal}{\emph{ACM Transactions on Knowledge Discovery from
  Data (TKDD)}} \bibinfo{volume}{13(4)}, \bibinfo{number}{37}
  (\bibinfo{year}{2019}).
\newblock


\bibitem[\protect\citeauthoryear{Yao, Torabi, Cho, Ballas, Pal, Larochelle, and
  Courville}{Yao et~al\mbox{.}}{2015}]%
        {yao2015video}
\bibfield{author}{\bibinfo{person}{Li Yao}, \bibinfo{person}{Atousa Torabi},
  \bibinfo{person}{Kyunghyun Cho}, \bibinfo{person}{Nicolas Ballas},
  \bibinfo{person}{Christopher Pal}, \bibinfo{person}{Hugo Larochelle}, {and}
  \bibinfo{person}{Aaron Courville}.} \bibinfo{year}{2015}\natexlab{}.
\newblock \showarticletitle{Video description generation incorporating
  spatio-temporal features and a soft-attention mechanism}.
\newblock \bibinfo{journal}{\emph{arXiv preprint arXiv:1502.08029}}
  (\bibinfo{year}{2015}).
\newblock


\bibitem[\protect\citeauthoryear{You, Jia, Liu, and Yang}{You
  et~al\mbox{.}}{2019}]%
        {you2019improving}
\bibfield{author}{\bibinfo{person}{Yongjian You}, \bibinfo{person}{Weijia Jia},
  \bibinfo{person}{Tianyi Liu}, {and} \bibinfo{person}{Wenmian Yang}.}
  \bibinfo{year}{2019}\natexlab{}.
\newblock \showarticletitle{Improving Abstractive Document Summarization with
  Salient Information Modeling}. In \bibinfo{booktitle}{\emph{Proceedings of
  the 57th Conference of the Association for Computational Linguistics}}.
  \bibinfo{pages}{2132--2141}.
\newblock


\end{thebibliography}

%

\end{document}